# A taxonomy of localization techniques based on multidimensional scaling

Biljana Risteska Stojkoska
Faculty of Computer Science and Engineering (FCSE)
University "Ss. Cyril and Methodius"
Skopje, Macedonia
biljana.stojkoska@finki.ukim.mk

*Abstract*—Localization in Wireless Sensor Networks (WSNs) has been a challenging problem in the last decade. The most explored approaches for this purpose are based on multidimensional scaling (MDS) technique. The first algorithm that introduced MDS for nodes localization in sensor networks is well known as MDS-MAP. Since its appearance in 2003, many variations of MDS-MAP have been proposed in the literature. This paper aims to provide a comprehensive survey of the localization techniques that are based on MDS. We classify MDS-based algorithms according to different taxonomy features and different evaluation metrics.

## I. INTRODUCTION

Wireless Sensor Networks (WSNs) have been a hot research area in the last decade. The potential application of WSN include different military and civilian application for monitoring (environmental, healthcare, structural health, smart home or smart grid)[1].

In the recent years, the concept of scattered devices dedicated to one-purpose application has been replaced with Internet of Things (IoT) and machine-to-machine (M2M) communication for general-purpose application [2]. However, the basic principles and challenges associated with WSNs remain attractive for the research communities. Moreover, with the extended range of potential WSN applications, these challenges are being reinvented and redesigned.

Localization in WSN refers to the process of discovering the locations of the nodes in the network. Many of the algorithms for WSN localization proposed in the literature are based on multidimensional scaling technique (MDS). This approach is very robust even if the network has only small portion of anchor nodes (nodes with a priori known location). Since 2003, when this approach was firstly presented on INFOCOM as MDS-MAP [3], new modifications and improvements were constantly developed by the researchers worldwide.

Multidimensional scaling (MDS) is a set of techniques used for reducing the dimensionality of the data (objects). MDS visualizes the results in order to show hidden structures in the data [4]. MDS algorithm uses the distances between each pair of object as an input and generates 2D-points or 3D-points as an output.

Multidimensional scaling as a technique for WSN localization consists of the following three steps:

- Step 1: Calculate the distances between every pair of nodes and generate a distance matrix that serves as an input to the step 2.

- Step 2: Apply multidimensional scaling (classical or non-metric) to the distance matrix. The first largest eigenvalues and eigenvectors give a relative map with relative location for each node.

- Step 3: Transform the relative map into absolute map using sufficient number of anchor nodes.

The basic pipeline of MDS algorithm for nodes localization is shown on Fig. 1. The input to the algorithm is available distance measurements between the nodes. The output can be obtained after step 2 as a relative map of the network, or after step 3 as an absolute map of the network.

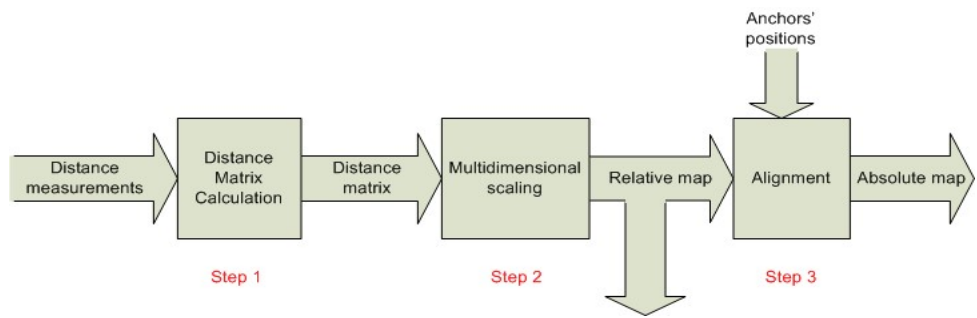

Fig. 1. Pipeline of MDS-based localization techniques.

In this paper, we present taxonomy of localization techniques based on multidimensional scaling algorithm. This taxonomy can be easily mapped to other localization approaches that use similar nonlinear dimensionality reduction techniques. For example, self organizing maps (SOMs) [5][6], manifold learning algorithm [7] or semidefinite programming [8][9] can follow same or similar taxonomy like the one presented in this paper. However, most of these approaches are represented with limited number of publications compared to MDS with respect to solving WSN localization problem.

The rest of this paper is organized as follows. In the next section, we present the taxonomy features used to distinguish different localization technique based on MDS. In the third section, we elaborate the metrics used for evaluation of the MDS-based techniques. Section four surveys different MDS techniques and qualitatively compares them regarding different taxonomy features and evaluation metrics. This paper is concluded in section five.

II. TAXONOMY FEATURES OF MDS-BASED TECHNIQUES

We identify five taxonomy features of localization techniques based on multidimensional scaling (Fig.2):

A. Computational model

WSNs are characterized with limited resources (CPU, memory, energy, etc.). Hence, choosing the most suitable computational model is not a trivial task. We define the computation model with respect to nodes' tasks in MDS computational pipeline. We indentified three main computational models.

*1) Centralized approach*

In centralized localization approaches, all distance measurements between neighboring nodes are collected at one central point (base station or sink). The sink node has three main tasks:

- to construct the distance matrix
- to apply MDS on the distance matrix
- to convert the relative map into absolute map (if anchors' positions are available)

which correspond to the three steps of the algorithm.

The main drawback of this model is its communicational overhead, as all distance measurements need to be propagated to the sink node. For a large scale networks, this requires retransmission through many intermediate nodes, which leads to additional energy consumption. On the other side, the step 2 of the algorithm will benefit from this model, as MDS will tend to minimize the overall positioning error.

*2) Fully distributed approach*

In distributed approaches, the distance measurements are not propagated to the sink node. Instead, they are broadcasted only in the near proximity. Each node is responsible to collect the distance measurements from its one-hop or two-hop neighbors. Then, the node uses these measurements to construct the distance matrix and to perform MDS in order to obtain the relative map of its own neighborhood. The relative maps from all nodes can later be propagated to the sink node, where they can be merged into one global map. In this approach, step 1 and step 2 are performed at the nodes' side, while step 3 (if applicable) is performed at the sink side.

As a special case of fully distributed approach is self-localization using location assistants (LA). Location-unaware nodes measure a set of distances to LAs, which are stationary or mobile nodes (e.g. airplane) with stronger radio signal. LAs broadcast their absolute location. Having radio signal strengths and LAs coordinates, location-unaware nodes perform MDS to obtain its own location [10].

*3) Cluster-based approach*

This approach is also known as hierarchical. Here, the network is divided into clusters which represent group of nodes in close proximity. One node in the cluster (known as cluster-head) is responsible to collect the distance measurements from all members of that cluster, to construct the distance matrix and to obtain relative map of its own cluster.

This approach is very similar with the fully distributed, except that only the cluster heads are responsible to run the algorithm. It is especially suitable for irregularly shaped networks (C-shape, O-shape, H-shape, etc.). If applied on networks with regular shape, this approach performs worse than centralized approach, due to cumulative error from the merging process.

B. Dimensionality of the network

We define three types of network regarding network dimensionality: two-dimensional (2D), three-dimensional (3D) and surface networks. Two-dimensional network is suitable for flat indoor and outdoor environments (flat agricultural field, one-floor apartment, etc.). Most of the outdoor networks fit into surface type (multi-floor shopping mall, hill, valley, etc.), while underwater networks can be considered as pure three-dimensional networks.

When applying MDS pipeline, step 2 and step 3 differ for different networks' dimensionality (Table 1).

TABLE I. NETWORK DIMENSIONALITY REGARDING MDS PIPELINE

| Network dimensionality | Number of eigenvalues (step 2) | Minimal number of anchors needed for alignment (step 3) |
|---|---|---|
| 2D | first 2 largest | 3 |
| 3D | first 3 largest | 4 |
| surface | first 3 largest | 4 |

C. Deployment environment

Regarding the environment where the nodes are deployed, we consider indoor and outdoor networks. The main difference between these types is availability of anchor nodes. In outdoor environment, nodes equipped with Global Positioning System (GPS) can become anchor nodes. Since GPS signals are not available in indoor environments, the role of anchors belongs to static nodes that are fixed with the buildings' infrastructure and their location is usually recorded manually. Underwater network is considered as outdoor; still, GPS signals are not available.

Other important difference is the radio propagation model, which is dependent on environmental condition (temperature, humidity, etc) and is highly variant in outdoor environment.

*D. Anchors*

Depending on the WSN application, in some cases there is no need to obtain the absolute location of the nodes. Users are interested in node location with respect to other nodes in the network. In such cases, step 3 of the pipeline is not needed.

If anchor nodes are available, the relative map can be easily translated into absolute map using geometric transformations (translation, rotation, and reflection). At least three (four) anchors are needed for 2D (3D) networks.

Some algorithms based on MDS allow random placement of the anchors, while other have more strict requirements about the anchors' position, i.e. allow only anchors placed at the edges of the deployment area.

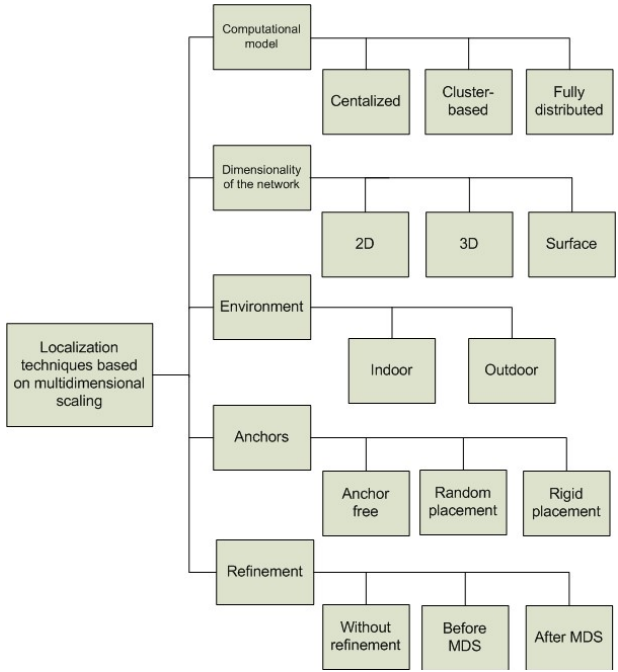

Fig. 2. Taxonomy of MDS-based localization techniques

*E. Refinement*

MDS is very accurate technique for dimensionality reduction. If the correct distance matrix is given as input, MDS algorithm will reconstruct the map of the network without error. But, calculating distance matrix for networks where only distances between neighboring nodes are known is not a trivial task. This problem in MDS-MAP is solved by applying Dijkstra's or Floyd's all pairs shortest path algorithm [11]. Dijkstra's algorithm is a graph search algorithm that solves the single-source shortest path problem. In WSN localization problem, the sensor network is represented as a graph with non-negative edge path costs, while the real,

Euclidean distance between two non-neighboring nodes is replaced with the distance calculated using Dijkstra (or other) algorithm. But the assumption that Dijkstra distance between two nodes correlates with their Euclidean distance is hardly true. This approximation produces an error, i.e., the positions obtained after alignments usually differ from the real positions. The error is bigger when the nodes are in multi-hop communication range, which is a common case in obstructed environments. It is usually caused by the presence of obstacles or terrain irregularities that can obstruct the line-of-sight between nodes or cause signal reflections.

Refinement process aims to reduce this error using different optimization techniques. Refinement can be applied in step 1 and in step 3 from the MDS pipeline:

*1) Refinement before MDS*

This refinement aims to minimize the inter-node distance error between non-neighboring nodes in the distance matrix. Variations of MDS-MAP use other approaches for distance matrics calculation, that are not based on Dijkstra's or Floyd's all pairs shortest path algorithm. Those techniques apply additional geometric constrants and produce better distance matrices compared to Dijkstra's or Floyd's algorithms.

*2) Refinement after MDS*

This refinement tends to minimize inter anchor distances error. The MDS-based techniques generate the coordinates of all nodes, including the anchor nodes. Moreover, after the alignment, the predicted positions for all nodes are obtained, including the anchor nodes. Although predicted coordinates of the anchors are usually discarded because their exact positions are known in advance, they can be used to refine other predicted positions. Through refinement, the predicted anchor positions are shifted toward real anchors positions. The refinement algorithms are usually iterative, and this shifting is actually performed for each node, toward a position for which the localization error is smaller.

### III. EVALUATION METRICS

In this section, we are describing different metrics used to evaluate the accuracy and the quality of the localization techniques.

*A. Localization error*

The main objective of localization is to predict the nodes location as accurate as possible. In order to evaluate the prediction error, different expressions of the error are adopted in the literature. The difference between the real and the predicted positions is known as estimation error, localization error or positioning error. For WSN that consists of W unknown nodes, where $(x_i, y_i, z_i)$ is the real position and $(x_i', y_i', z_i')$ is the predicted position of i-th node, the localization error can be expressed as:

*1) Average localization error(ALE)*

$$ALE = \frac{1}{W} \sum_{i=1}^{W} \sqrt{(x_i - x_i')^2 + (y_i - y_i')^2 + (z_i - z_i')^2} \quad (1)$$

*2) Root-mean-square error (RMSE)*

$$RMSE = \sqrt{\frac{1}{W}\sum_{i=1}^{W}(x_i - x_i')^2 + (y_i - y_i')^2 + (z_i - z_i')^2} \quad (2)$$

Localization techniques can be evaluated by comparing the localization accuracy obtained using (1) or (2) with the corresponding Cramer-Rao lower bound (CRLB), which is a lower bound on the variance of the estimator for locations. CRLB is used to assess the optimum achievable localization accuracy which can be attained with the available measurement set. CRLB is computed using the Likelihood function in which the covariance of the measurement vector enters as a factor [12].

*B. Radio range error*

Different techniques can be used for inter-node distance measurements, like Radio Signal Strength Indicator (RSSI), Time of Arrival (ToA) or Time Difference of Arrival (TDoA) [13].

Among them, RSSI is the most popular since it doesn't require any additional hardware. On the other side, RSSI is not that accurate, due to the different reasons, like different propagation model or device calibration. When modeling the radio range, we need to consider radio range error which is at least 20%, but more often is greater [13].

*C. Communication cost*

WSN has limited energy resources. It is known that data transmission spends much more energy compared with data processing. Hence, it is very important in WSN to reduce the number of transmissions and to apply local processing whenever possible.
Centralized approaches for WSN localization require all distance measurements to be transmitted to the sink node. In multi-hop network, this requirement can lead to battery drain.

*D. Computation cost*

Although local computations are more preferable for energy saving, they are not always fusible due to CPU limitations. WSN nodes are usually equipped with limited CPU and memory capacities, which cannot handle exhaustive computations, or at least not in expected time span.

Considering communication cost and computation cost, the ideal algorithm for localization should maintain a tradeoff between energy saving and required localization accuracy.

*E. Network density*

MDS-based techniques require connected network, which means that there is a path between every pair of nodes in the network. Moreover, greater network density leads to better localization accuracy. This is rather expected knowing that the main factor that affects the localization performance is the distance matrix. More one-hop neighbors will lead to more accurate distance matrix, which will decrease the overall localization error. The most common parameter used to represent the network density is network connectivity, which is the average number of one-hop neighbors. Therefore, radio range is a crucial parameter that directly affects network connectivity. By extending the radio range, one can increase the network connectivity, or the network density.

However, a dense network does not necessarily guarantee high accuracy, especially in networks with irregular topologies.

*F. Anchor location and anchor density*

Number of anchors and their location affect the localization accuracy. As expected, more anchor nodes will improve the performance of the MDS-based technique, but increasing the number of anchors does not have a crucial influence on the localization error. This is especially notable for networks with high connectivity levels.

*G. Irregular vs regular network topology*

Most of the techniques in the literature are evaluated on networks with regular topologies (evenly distributed nodes and anchors). To achieve this regularity, simulations are usually performed on grid, hex or random topology, although the later can barely be considered as evenly distributed.

However, real wireless networks have irregular shape due to presence of obstacles in the environment where WSN is deployed. In order to obtain the performances of the algorithm, simulations and real experiment should also consider irregular network topologies. Most simulation scenarios consider C-shape or H-shape networks to investigate the performances for irregular networks.

Localization error in the case of networks with irregular topology is generally greater than in the case of regular network topologies. Better results can be obtained if network is divided into sub-segments, which can be considered as sub-networks with regular topology.

IV. QUALITATIVE COMPARISON OF MDS-BASED TECHNIQUES

In this section, we are going to describe the most important MDS-based techniques for nodes localization in WSN. Among dozens, we have chosen the most representative, which were inspiration for new improvements. Table II provides a qualitative comparison of these techniques regarding different taxonomy features, localization error and computation cost. The later is denoted as *O(n)*, where *n* is the total number of nodes in the network.

Hereafter, we are going to present them in chronological order.

MDS-MAP is the first approach based on MDS [3]. It follows the basic MDS pipeline without any refinements. In step 1, MDS-MAP uses Dijkstra algorithm for distance matrix calculation.

MDS-MAP(P, R) is a distributed localization algorithm, where each node creates a local map within its two-hop neighbors using classical MDS algorithm [14]. Each local map is then refined with least-squares minimization.

MDS-MAP(P, O) is distributed algorithm and can be considered as an extension of MDS-MAP(P, R). The modification is the use of the ordinal MDS (instead of classical MDS) during the estimation phase [15].

TABLE II. QUALITATIVE COMPARISION OF DIFFERENT MDS BASED TECHNIGUES

| Author | Year | Techniques' acronym | Taxonomy feature | | | | | Evaluation metric | |
|---|---|---|---|---|---|---|---|---|---|
| | | | Computational model | Anchor placement | Dimensionalty | Refinement before MDS | Refinement after the alignment | Localization Error | $O(n)$ |
| Shang [3] | 2003 | MDS-MAP | Centralized | random | 2D | NA | NA | ALE | $O(n^3)$ |
| Shang [14] | 2004 | MDS-MAP (P,R) | Fully distributed | random | 2D | NA | least squares | ALE | $O(n)$ |
| Vivekanandan [15] | 2006 | MDS-MAP (P, O) | Fully distributed | random | 2D | iterative monotone regression | least squares | ALE | $O(n)$ |
| Stojkoska [16] | 2008 | CB-MDS | Cluster-based | random | 2D | NA | NA | ALE | $O(n)$ |
| Chaurasiya [17] | 2013 | NDEA | Centralized | random | 3D | iterative distance estimation | NA | ALE, RMSE | $O(n^3)$ |
| Stojkoska [18][19][20] | 2013 | IMDS | Centralized | random | 2D, 3D, surface | heuristic approach based on averaging | NA | ALE | $O(n^3)$ |
| Saeed [21] | 2014 | MDS-LM | Centralized | random | 2D | NA | Levenberg–Marquardt | RMSE, CRLB | $O(n^3)$ |
| Stojkoska & Saeed [22] | 2015 | MHL-M | Centralized | random | 3D | heuristic approach based on averaging | Levenberg–Marquardt | RMSE, CRLB | $O(n^3)$ |
| Fan [23] | 2015 | D3D-MDS | Cluster-based | edge | 3D | NA | least squares | ALE | $O(n)$ |

CB-MDS is a cluster-based approach where MDS is performed at cluster level [16]. At the end, all cluster maps are merged in one global relative map, which is centrally align into absolute map. This approach has very acceptable error on the case of networks with irregular topologies.

HDEA is a novel distance estimation approach [17] that uses iterative algorithm for distance matrix calculation. In order to estimate the distance between two nodes, this approach uses group of three common neighboring nodes. This requires high network connectivity in order to produce accurate inter node distance estimation.

IMDS is an improved version of MDS-MAP, where Dijkstra algorithm from step 1 is replaced with heuristic algorithm for inter-node distance estimation based on averaging [18-20]. This algorithm has implementations for 2D, 3D and surface networks.

MDS-LM is a centralized approach that uses Levenberg–Marquardt optimization in step 3 [21].

MHL-M [22] combines IMDS [20] and MDS-LM [21]. Computational cost is the same as IMDS, but the accuracy is much better. For high connectivity, this approach achieves Cramer-Rao Lower Bound (CRLB).

D3D-MDS [23] is a 3D version of CB-MDS [16]. Hence, its benefits are not sufficiently explored on networks with irregular topologies. For regular networks, IMDS [18-20] performances are much better than D3D-MDS, which is rather expected. Cluster-based approaches should be favored only for irregular topologies.

Most of the algorithms based on MDS generally assume that RSSI is used to generate the measurement set. However, there are many approaches in the literature that assume angle of arrival (AoA) measurement set. For example, in [24], the authors propose an algorithm (RAST) that is very similar to MDS-MAP, but instead of MDS uses Singular Value Decomposition (SVD) approach for dimensionality reduction. Cluster-based solution of RAST is proposed in [25]. These approaches follow the taxonomy proposed in this paper. Regarding the evaluation metrics, instead of radio range error, these approaches should consider standard deviation of AoA measurements.

Approaches overviewed in this paper are evaluated only through simulations. In order to provide an exact quantitative comparison of these algorithms, they need to be re-implemented and evaluated under same simulation setup. In the simulation scenarios, algorithms are evaluated considering large scale networks, from 100 up to 1000 wireless nodes. Therefore, there is lack of experimental evaluation of these algorithms in the literature, and, if there are any, they are performed on small scale networks, where nodes are in close proximity to each other. With the development of new technologies, problem of WSN localization has overlapped with the problem of smartphones' localization, especially in indoor environments, where GPS signals are not available. In the near future, it is expected that new crowdsensing and/or crowdsourcing approaches can be used to experimentally verify most of the algorithms for localization, as well as those based on MDS.

V. CONCLUSION

Multidimensional scaling is a very explored concept for localization in Wireless Sensor Networks. Although initially proposed more than ten years ago, it is still attractive among researchers.

This paper is a valuable introduction for the newcomers to the field of localization in WSN, especially for those that aim to investigate and improve MDS-based techniques.

It can also serve as a tutorial for those that are implementing solutions for localization, in order to help them choose the most suitable MDS approach for their particular problem regarding different network characteristics and hardware limitations.

REFERENCES

[1] B. Risteska Stojkoska, A. Popovska Avramova, and P. Chatzimisios, "Application of wireless sensor networks for indoor temperature regulation," *International Journal of Distributed Sensor Networks* 2014 (2014).

[2] L. Mainetti, L. Patrono, and A. Vilei, "Evolution of wireless sensor networks towards the internet of things: A survey," In *Software, Telecommunications and Computer Networks (SoftCOM), 2011 19th International Conference on*, pp. 1-6.

[3] Y. Shang, W. Ruml, Y. Zhang & P. J. M. Fromherz, Localization from mere connectivity, in Proceedings of the 4th ACM International Symposiumon Mobile Ad Hoc Networking, Annapolis, Maryland, USA, pp. 201–212, (2003).

[4] T. Cox and M. Cox, Multidimensional Scaling, Chapman & Hall, London, UK, 1994.

[5] G. Giorgetti, S. K. S. Gupta, and G. Manes, "Wireless localization using self-organizing maps," in Proceedings of the 6th International Conference on Information Processing in Sensor Networks, ser. IPSN '07. New York, NY, USA: ACM, 2007, pp. 293–302.

[6] J. Hu and G. Lee, "Distributed localization of wireless sensor networks using self-organizing maps," in IEEE International Conference on Multisensor Fusion and Integration for Intelligent Systems, 2008, pp. 284–289.

[7] N. Patwari and A. O. Hero III. "Manifold learning algorithms for localization in wireless sensor networks." *Acoustics, Speech, and Signal Processing, 2004. Proceedings.(ICASSP'04). IEEE International Conference on*. Vol. 3. IEEE, 2004.

[8] P. Biswas and Y. Ye. "Semidefinite programming for ad hoc wireless sensor network localization." *Proceedings of the 3rd international symposium on Information processing in sensor networks*. ACM, 2004.

[9] B. Stojkoska, I. Ivanoska, and D. Davcev. "Wireless sensor networks localization methods: Multidimensional scaling vs. semidefinite programming approach." *ICT Innovations 2009*. Springer Berlin Heidelberg, 2010. 145-155.

[10] X. Zhou, L. Zhang, and Q. Cheng, "Landscape-3D: a robust localization scheme for sensor networks over complex 3D terrains," in Proceedings of the 31st Annual IEEE Conference on Local Computer Networks (LCN '06), pp. 239–246.

[11] R. W. Floyd, "Algorithm 97: shortest path," Communications of the ACM, vol. 5, no. 6, p. 345, 1962.

[12] K. Regina, J. Hörst, and W. Koch. "Accuracy analysis for TDOA localization in sensor networks." *Information Fusion (FUSION), 2011 Proceedings of the 14th International Conference on*. IEEE, 2011.

[13] J. Zhang and L. Zhang, "Research on distance measurement based on RSSI of ZigBee," in ISECS International Colloquium on Computing, Communication, Control, and Management (CCCM '09), vol. 3, pp. 210–212.

[14] Y. Shang, W. Ruml, Improved MDS-based localization, in Proceedings of the IEEE INFOCOM Twenty-third Annual Joint Conference of the Computer and Communications society, vol.4, (2004).

[15] V.Vivekanandan, and V.W.Wong, 2006. Ordinal MDS-based localisation for wireless sensor networks. *International Journal of Sensor Networks*, 1(3-4), pp.169-178.

[16] B.Stojkoska, D. Davcev, and A. Kulakov. "Cluster-based MDS algorithm for nodes localization in wireless sensor networks with irregular topologies," In *Proceedings of the 5th international conference on Soft computing as transdisciplinary science and technology*, ACM, 2008, pp. 384-389.

[17] V.K.Chaurasiya, N.Jain, and G.C.Nandi, 2014. A novel distance estimation approach for 3D localization in wireless sensor network using multi dimensional scaling. *Information Fusion*, 15, pp.5-18.

[18] B. Risteska Stojkoska, D. Davcev, MDS-based Algorithm for Nodes Localization in 3D Surface Sensor Networks. In Proceedings of the 7th International Conference on Sensor Technologies and Applications (SENSORCOMM'13) 2013, pp. 44-50.

[19] B. Risteska Stojkoska and V. Kirandziska. "Improved MDS-based algorithm for nodes localization in wireless sensor networks," In *EUROCON, 2013 IEEE*, pp. 608-613.

[20] B. Risteska Stojkoska, Nodes Localization in 3D Wireless Sensor Networks Based on Multidimensional Scaling Algorithm, *International Scholarly Research Notices* 2014 (2014).

[21] N. Saeed & H. Nam, MDS-LM for Wireless Sensor Networks Localization, in Proceedings of the IEEE 79th Vehicular Technology Conference, Seoul (Korea), pp. 1–6, (May 2014).

[22] N. Saeed, and B. Risteska Stojkoska, 'Robust localisation algorithm for large scale 3D wireless sensor networks', to appear in Int. J. Ad Hoc and Ubiquitous Computing.

[23] J. Fan, B. Zhang, and G. Dai, "D3D-MDS: A Distributed 3D Localization Scheme for an Irregular Wireless Sensor Network Using Multidimensional Scaling," International Journal of Distributed Sensor Networks, vol. 2015.

[24] J.N. Ash, and L. C. Potter. "Robust system multiangulation using subspace methods." *Proceedings of the 6th international conference on Information processing in sensor networks*. ACM, 2007.

[25] A. Damir. "Distributed algorithm for anchor-free network localization using angle of arrival." *Industrial Electronics, 2008. ISIE 2008. IEEE International Symposium on*. IEEE, 2008.